\documentclass[11pt]{article}
\input epsf.sty  
\def \lket {|}
\def \rket {\rangle}

\def\bbbz{Z}

\def \H {{\cal H}}
\newcommand{\ket}[1]{\lket #1\rket}

\newcommand{\comment}[1]{}
\newtheorem{theorem}{Theorem}
\begin{document}

\title{Quantum walks and their algorithmic applications}

\author{Andris Ambainis\footnote{Computer 
Science Division, University of California,
Berkeley, CA 94720, USA,
ambainis@cs.berkeley.edu}}

\maketitle

\begin{abstract}
Quantum walks are quantum counterparts of Markov chains.
In this article, we give a brief overview of quantum walks,
with emphasis on their algorithmic applications.
\end{abstract}

\section{Introduction}

Markov chains have an important role in randomized algorithms 
(for example, algorithms for approximating permanent and other 
hard problems and volume estimation \cite{MR}).
Recently, quantum counterparts of Markov chains have been 
defined \cite{AZ,Meyer,AA+,AB+}, 
with the hope that they could have similar applications
in design of new quantum algorithms.
It was immediately noticed that quantum walks behave
quite differently from their classical counterparts.
Recently, several algorithms using quantum walks have
been invented.

In this paper, we survey results on algorithmic applications of
quantum walks. Sections \ref{sec:line} and \ref{sec:gen}
give the background on quantum walks. In particular, section \ref{sec:line}
explains basics of quantum walks, using quantum walk on the line
as an example. Then, in section \ref{sec:gen} we show how to 
define a quantum walk on an arbitrary graph. 
After that, in sections \ref{sub:hit} and \ref{sub:qws}, 
we describe the known algorithms
that use quantum walks. Those algorithms can be classified
into two groups: algorithms achieving exponentially faster hitting 
times \cite{FG,CFG,CC+,KempeHit} and quantum walk search 
\cite{Shenvi,CG,ADistinct,AKRS}.
We describe the first group in section \ref{sub:hit} and the second
group in section \ref{sub:qws}.
At the conclusion, we present some open problems in
section \ref{sec:open}.

Besides the algorithmic aspects described in the current paper,
there are many other interesting results about quantum walk.
We refer the reader to an excellent survey by Kempe \cite{Kempe}
for more information about quantum walks in general.

\section{Quantum walk on the line}
\label{sec:line}

Quantum walk on a line is a simple example which shows
many properties of quantum walks. 
It is also often useful as a tool in the analysis 
of quantum walks on more complicated graphs.
(We will show two examples of that in sections \ref{sub:hit}
and \ref{sub:qws}.)

\subsection{Discrete quantum walk}
\label{sec:disc}

In a classical random walk on the line, we start in
location 0. At each time step, we move left with
probability 1/2 and right with probability 1/2.

The straightforward counterpart would be a quantum process
with basis states $\ket{n}$, $n\in\bbbz$. At each
time step, it would perform a unitary transformation
\begin{equation}
\label{eq:non} 
\ket{n}\rightarrow a\ket{n-1}+b\ket{n}+c\ket{n+1}, 
\end{equation}
which means move left with some amplitude $a$,
staying at the same place with amplitude $b$ and moving
right with amplitude $c$. Moreover, we would like
the process to behave in the same way in every location.
That is, $a$, $b$ and $c$ should be independent of $n$
(just like the probabilities of moving left/right are
independent of $n$ in the classical walk).
Unfortunately, this definition does not work.

\begin{theorem}
\cite{Meyer}
The transformation $U$ defined by equation (\ref{eq:non})
is unitary if and only if one of following three conditions is true:
\begin{enumerate}
\item
$|a|=1$, $b=c=0$;
\item
$|b|=1$, $a=c=0$;
\item
$|c|=1$, $a=b=0$;
\end{enumerate}
\end{theorem}

Thus, the only possible transformations are the trivial ones (ones
that, at each step, either always move left or always stay in place
or always move right).
The same problem also appears when defining quantum walks on
many other graphs.

It can be solved by introducing an additional ``coin'' state.
We consider the state space consisting of states $\ket{n, 0}$
and $\ket{n, 1}$ for $n\in\bbbz$. At each step, we do 
two operations:
\begin{enumerate}
\item
A ``coin flip transformation'' $C$
\[ C\ket{n, 0} = a\ket{n,0}+b\ket{n,1} ,\]
\[ C\ket{n, 1} = c\ket{n,0}+d\ket{n,1} .\]
\item
Shift $S$:
\[ S\ket{n,0}=\ket{n-1,0} , \mbox{~~} S\ket{n,1}=\ket{n+1,1} .\]
\end{enumerate}
A step of quantum walk is $SC$.
The most common choice for $C$ is the Hadamard transformation
\begin{equation}
\label{eq:cf} 
\left( \begin{array}{cc}
a & b \\ c & d \end{array} \right) =
\left( \begin{array}{cc}
\frac{1}{\sqrt{2}} & \frac{1}{\sqrt{2}} \\ \frac{1}{\sqrt{2}} 
& -\frac{1}{\sqrt{2}} \end{array} \right) .
\end{equation}
Any other unitary transformation in 2-dimensions 
can be also used.

We can think of $C$ as a quantum counterpart of coin flip
in which we decide in which direction to move.
To see this analogy, consider the modification in which, 
between $C$ and $S$, we measure the state.
If the state before $C$ is $\ket{n, 0}$, then the state after 
$C$ is $\frac{1}{\sqrt{2}}\ket{n, 0}+\frac{1}{\sqrt{2}}\ket{n, 1}$
and measuring it yields $\ket{n, 0}$ and $\ket{n, 1}$
with probabilities 1/2 each. 
If the state before $C$ is $\ket{n, 1}$, then the state after 
$C$ is $\frac{1}{\sqrt{2}}\ket{n, 0}-\frac{1}{\sqrt{2}}\ket{n, 1}$.
Measuring this state also yields $\ket{n, 0}$ and $\ket{n, 1}$
with probabilities 1/2 each. 
Thus, $C$ is now equivalent to probabilistically
picking one of $\ket{n, 0}$ and $\ket{n, 1}$
with probabilities 1/2 each. 

However, in the quantum coin flip we do not measure the state.
This leads to a clearly different results. 
Consider the first three steps of quantum walk, starting in state 
$\ket{0, 0}$.
\[ \ket{0, 0}\rightarrow 
\frac{1}{\sqrt{2}} \ket{0,0}+\frac{1}{\sqrt{2}} \ket{0,1}\rightarrow 
\frac{1}{\sqrt{2}} \ket{-1,0}+\frac{1}{\sqrt{2}} \ket{1,1}\rightarrow \]
\[ \frac{1}{2} \ket{-1,0} + \frac{1}{2} \ket{-1,1}+
\frac{1}{2} \ket{1,0} - \frac{1}{2} \ket{1,1} \rightarrow \]
\[ \frac{1}{2} \ket{-2,0} + \frac{1}{2} \ket{0,1}+
\frac{1}{2} \ket{0,0} - \frac{1}{2} \ket{2,1} \rightarrow \]
\[ \frac{1}{2\sqrt{2}} \ket{-2, 0} + \frac{1}{2\sqrt{2}} \ket{-2, 1}+
\frac{1}{\sqrt{2}}\ket{0,0} - \frac{1}{2\sqrt{2}} \ket{2, 0}+
\frac{1}{2\sqrt{2}} \ket{2, 1} \rightarrow \]
\[ \frac{1}{2\sqrt{2}} \ket{-3, 0} + \frac{1}{2\sqrt{2}} \ket{-1, 1}+
\frac{1}{\sqrt{2}}\ket{-1,0} - \frac{1}{2\sqrt{2}} \ket{1, 0}+
\frac{1}{2\sqrt{2}} \ket{3, 1} .\]
The result of first two steps is quite similar to classical random walk.
If, after the second step, we measured the state,
we would find $n=-2$ and $n=2$ with probability 1/4 and $n=0$ 
with probability 1/2.
This is exactly what we would have had in a classical random walk.

The third step is different. In the coin flip stage, 
the state $\frac{1}{2} \ket{0,1}+\frac{1}{2} \ket{0,0}$ gets
mapped to $\frac{1}{\sqrt{2}} \ket{0, 0}$. In the classical random walk,
we would choose left and right direction with probabilities 1/2 each.
In the quantum case, we end up going to the left
with all the amplitude.
This is the result of quantum interference.

\begin{figure}[th]

 \begin{minipage}{0.5\linewidth}
\epsfxsize=3in
\hspace{0in}
\epsfbox{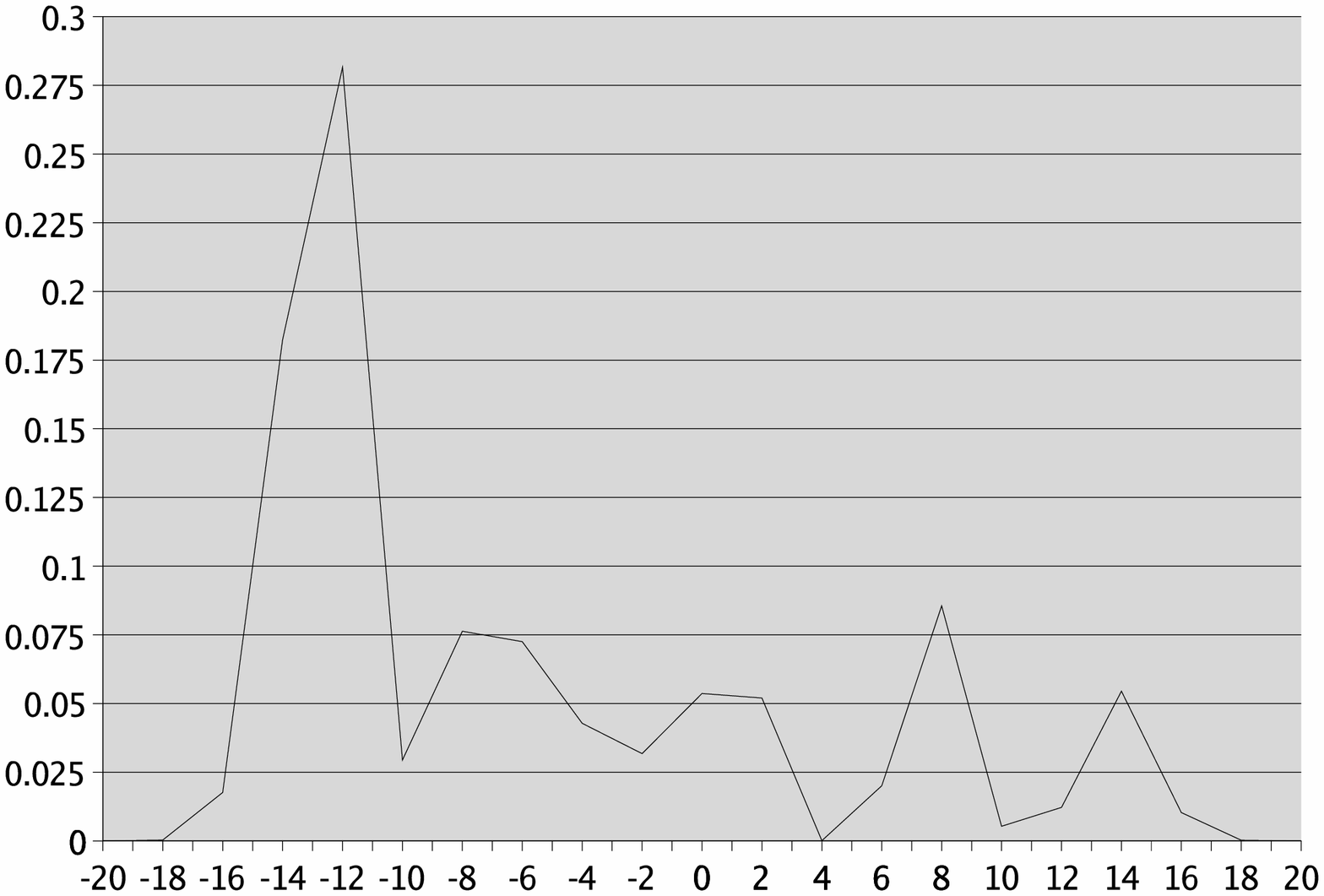}
\vspace*{8pt}
\caption{Probability distribution of quantum walk after 20 steps
with coin flip (\ref{eq:cf}).}
\label{fig:20}
\end{minipage}%
 \begin{minipage}{0.5\linewidth}
\epsfxsize=3in
\hspace{0in}
\epsfbox{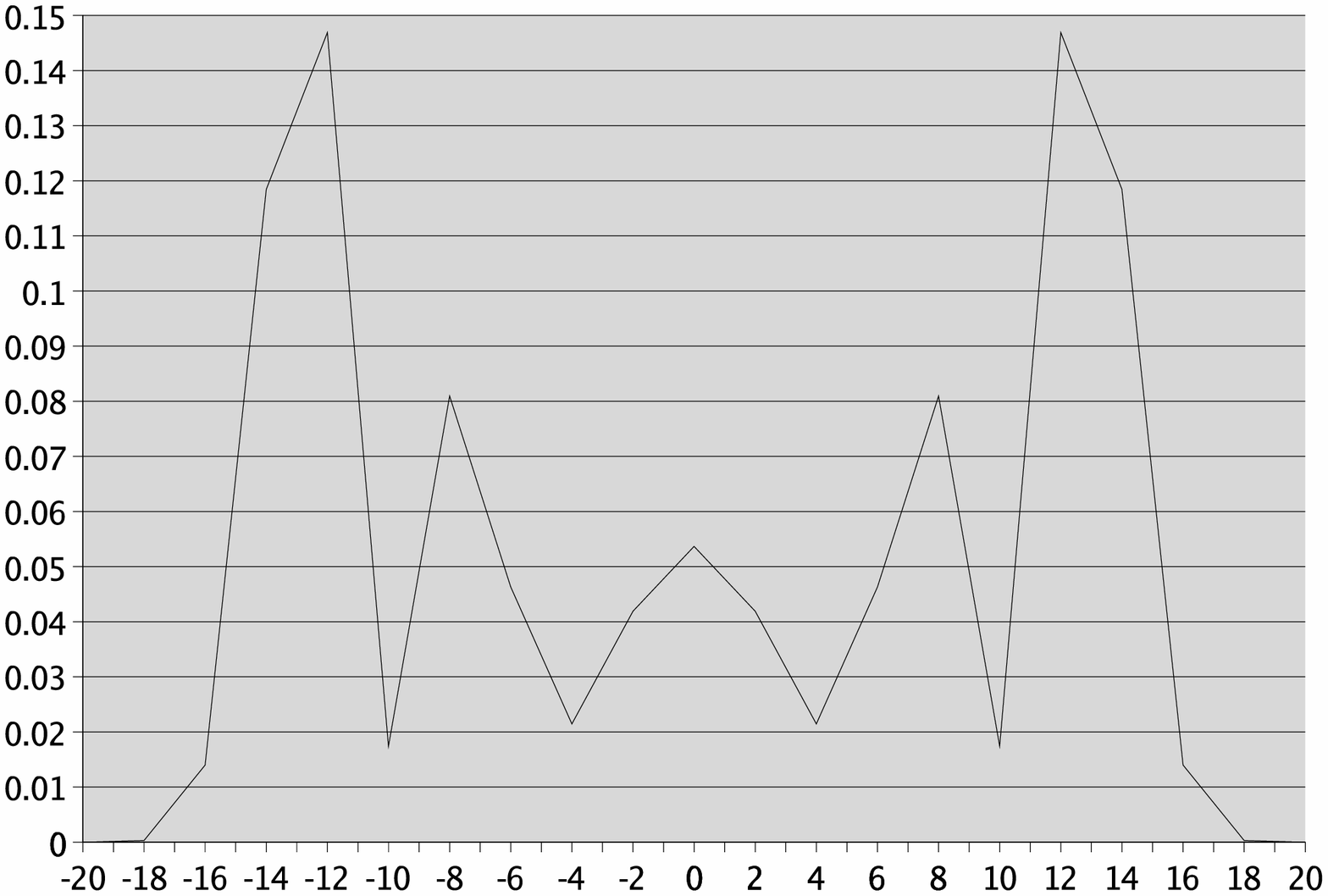}
\vspace*{8pt}
\caption{Probability distribution of quantum walk after 20 steps
with coin flip (\ref{eq:cfs}).}
\label{fig:20s}
\end{minipage}
\end{figure}

The difference becomes more striking as large times are considered. 
If we start at location 0 and run classical random walk for
$t$ steps, the probability distribution approaches normal distribution.
In the quantum case, the corresponding experiment would be
starting quantum walk in the state $\ket{0,0}$, 
running $t$ steps of quantum walk (which result in 
the transformation $(SC)^t$) and measuring the final state.
This yields probability distribution of the form shown in figure
\ref{fig:20} which is very different from the normal distribution.

\comment{
In figure \ref{fig:20}, we show the probability distributions generated by
classical random walk and quantum random walk in 20 steps. 
(For quantum walk, the distribution is generated by starting
in the state $\ket{0,0}$, performing $(SC)^t$ and measuring the
final state.)}

First, the distribution is biased towards left.
The drift to the left is a consequence of slightly non-symmetric
coin flip transformation. (In equation (\ref{eq:cf}), the amplitude
of going from $\ket{n,0}$ to $\ket{n,0}$ is $a=\frac{1}{\sqrt{2}}$
but the amplitude of going from $\ket{n,1}$ to $\ket{n,1}$ is 
$d=-\frac{1}{\sqrt{2}}$.) It disappears if more symmetric coin
\begin{equation}
\label{eq:cfs} 
\left(\begin{array}{cc} \frac{1}{\sqrt{2}} & \frac{i}{\sqrt{2}} \\
\frac{i}{\sqrt{2}} & \frac{1}{\sqrt{2}} \end{array} \right) 
\end{equation}
is used. The probability distribution for this coin transformation
is shown in figure \ref{fig:20s}. 
Two other quantum effects remain for almost any choice
of coin transformation and starting state\cite{AB+,NV}.
\begin{enumerate}
\item
The maximum probability is reached
for $|n|\approx \pm\frac{t}{\sqrt{2}}$. 
\item
There are $\Omega(t)$ locations $n$ for which 
the probability of measuring $\ket{n,0}$ or $\ket{n,1}$
is $\Omega(\frac{1}{t})$. This means that
the expected distance between the starting location $0$ and
the location which we measure after $t$ steps is $\Omega(t)$.
\end{enumerate}
Both of those sharply contrast with the classical random walk.
In the classical random walk, the maximum probability is reached 
for $n\approx 0$. This has a simple intuitive explanation:
the number of steps to the left is approximately equal to
the number of steps to the right. Second, the 
probability of being more than $O(\sqrt{t})$ steps
from the origin at time $t$ is negligible.

Can we exploit any of those properties to design fast quantum
algorithms? The second property seems to be most promising.
It means that quantum walk spreads quadratically faster
than its classical counterpart. 

The probability distribution generated by running quantum walk for
$t$ steps and measuring it has been studied in more detail in 
refs. \cite{AB+,NV,Line2,CS+,Line1}.
Other interesting quantum effects appear if we add boundary conditions
to the line \cite{AB+,BC+,Yamasaki}.

\subsection{Continous walk}
\label{sec:cont}

Farhi and Gutman \cite{FG} have defined continous time quantum walk.
In continous time, introducing ``coin'' state is not
necessary. Instead, one can define walk on the line on the set
of states $\ket{n}$, $n\in\bbbz$. To do that, we take Hamiltonian
$H$ defined by $H\ket{n}=-\ket{n-1}+2\ket{n}-\ket{n+1}$.
Running quantum walk for time $t$ is just applying the transformation 
$U^{iHt}$. 

The results about discrete and continous walk are often 
similar but the relation between the two is unclear.
In the classical world, continous walk can be obtained as a limit
of the discrete walk. In quantum case, this does not seem to be
true because discrete walk has a coin but continous walk does not.

\section{Quantum walks on general graphs}
\label{sec:gen}

How do we generalize the quantum walk on the line to more general graphs?
Continous walk can be extended to a general graph quite easily \cite{CFG}.
We take Hamiltonian $H$ defined by $H_{ij}=-1$ if $i\neq j$ and vertices $i$ and
$j$ are connected by an edge and $H_{ii}=d_i$ where $d_i$ is the degree of
vertex $i$. Then, running quantum walk for time $t$ is just applying
the transformation $e^{iHt}$.

For discrete quantum walk, defining it on an arbitrary graph is more
difficult. The first papers \cite{AA+} only defined it on graphs
where every vertex has the same number of outgoing edges.
However, there is a simple way to define a quantum walk on an arbitrary
undirected graph\footnote{This definition was first discovered by 
Watrous \cite{Wat} and then rediscovered by several other authors.
A different approach to this problem can be found 
in  \cite{Kendon}.}.

Let $V$ be the set of vertices and $E$ be the set of edges.
We use basis states $\ket{v, e}$ for all $v\in V$ and $e\in E$ such that
the edge $e$ is incident to the vertex $v$. 
One step of quantum walk now consists of the following:
\begin{enumerate}
\item
For each $v$, we perform a coin flip transformation $C_v$ on the states $\ket{v,e}$,
$e$ adjacent to $v$.
\item
We perform shift $S$ defined as follows. If edge $e$ has endpoints $v$ and $v'$,
then $S\ket{v, e}=\ket{v',e}$ and $S\ket{v',e}=\ket{v,e}$. 
\end{enumerate}
A very natural choice for $C_v$ is Grover's diffusion (first used in
Grover's search algorithm\cite{Grover}).
If there are $m$ edges $e$ incident to vertex $v$, we apply the transformation
described by matrix 
\[ D_m=\left( \begin{array}{cccc}
-1+\frac{2}{m} & \frac{2}{m} & \ldots & \frac{2}{m} \\
\frac{2}{m} & -1+\frac{2}{m} & \ldots & \frac{2}{m} \\
\ldots & \ldots & \ldots & \ldots \\
\frac{2}{m} & \frac{2}{m} & \ldots & -1+\frac{2}{m} 
\end{array} \right) \]
to states $\ket{v, e}$.
The logic behind this transformation is very simple.
Typically, we would like to treat all edges adjacent to $v$
in the same way. Thus, the amplitude of staying in the same state
$\ket{v,e}$ should be the same for all $e$. Similarly, for 
all $e$ and $e'$, $e\neq e'$, the amplitude of moving from
$\ket{v,e}$ to $\ket{v,e'}$ should be the same for all $e$ and $e'$.
The transformation $D_m$ satisfies these requirements.
Moreover, the only unitary matrices with real entries
satisfying this requirement are $D_m$, $-D_{m}$, $I$ (identity) and
$-I$. Since $I$ leads to trivial results (if we go from $v'$ to $v$
via edge $e$ in one step, we go back to $v'$ via the same edge in
the next step), $D_m$ is the natural choice.

A possible exception is if $v$ has exactly two edges adjacent to it.
Then, $D_m$ is equal to the matrix.
 \[ \sigma_x= \left( \begin{array}{cc}
0 & 1 \\
1 & 0
\end{array} \right) \]
A possible alternative in this case is the matrix
of equation (\ref{eq:cfs}) 
\comment{\[ \left( \begin{array}{cc}
\frac{1}{\sqrt{2}} & \frac{i}{\sqrt{2}} \\
\frac{i}{\sqrt{2}} & \frac{1}{\sqrt{2}}
\end{array} \right) \]}
which is complex valued but treats both of adjacent vertices in the same way.

Next, we describe the known quantum algorithms using quantum walks.
They can be broadly classified into two groups.

\section{Exponentially faster hitting}
\label{sub:hit}

\begin{figure}[th]
\epsfxsize=3in
\hspace{0in}
\epsfbox{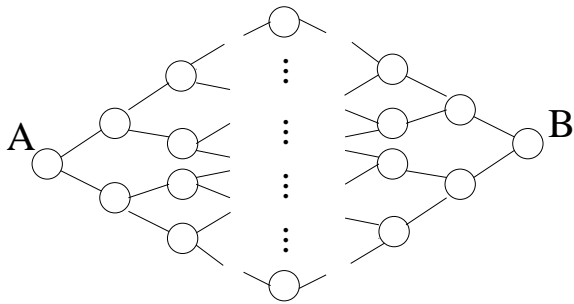}
\vspace*{8pt}
\caption{``Glued trees'' graph of Childs, Farhi and Gutman [7].}
\label{fig:childs}
\end{figure}

The algorithms of this type include  \cite{FG,CFG,CC+,KempeHit}.
In this group of algorithms, we have a known graph $G$.
We start in one vertex $A$ and would like to reach a certain target vertex $B$.
For example, in  \cite{CFG}, we have the graph shown in figure \ref{fig:childs}
and would like to go from the left vertex $A$ to the right vertex $B$.
At one step, we can choose one of edges adjacent to the current vertex
and move to the other endpoint of this vertex. The difficulty is that edges
and vertices are not labelled and we do not know which of possible moves
takes us in the right direction.

One solution would be to do classical random walk, starting from $A$.
Then, the number of steps till we first enter $B$ is typically 
polynomial in the 
number of vertices of $G$. 
For graphs with regular structure, 
quantum walks can do much better. For example\cite{CFG}, consider
two full binary trees of depth $d$ with roots $A$ and $B$.
Glue each leaf of the first tree with a leaf of the second tree.
The resulting graph is shown in figure \ref{fig:childs}.
It has $O(2^d)$ vertices. If we start in $A$ and do a classical
random walk, it will take $\Omega(2^d)$ steps to reach $B$.

Childs et al. \cite{CFG} showed that a continous\footnote{Computer
simulations suggest that a similar result is also true for
discrete walks, for example, with the definition of discrete 
walk on general graphs described in section \ref{sec:gen}.} quantum walk on $G$
can reach $B$ from $A$ in $O(d^2)$ steps.
They take basis states $\ket{i}$ with $i$ corresponding to vertices of $G$
and define a Hamiltonian $H$ by $H_{ij}=-1$ if $i, j$ are
adjacent and $H_{ii}=d_i$ (as described in section \ref{sec:gen}).
Then, $t$ steps of quantum walk is just the unitary transformation
$E^{iHt}$. 

It might seem that analyzing quantum walk on $G$ is more
difficult than analyzing the walk on line (because $G$
has a more complicated structure).
However, there is a clever reduction to the walk on line \cite{CFG}.
Partition vertices into $2d+1$ sets $S_{-d}$, $S_{-d+1}$,
$\ldots$, $S_d$. The set $S_{-d}$ contains just the vertex $A$.
The set $S_{-d+i}$ contains vertices which are reached by walking
$i$ edges to the right from $A$. In particular, $S_0$ contains all
the vertices of degree 2 in the middle and $S_d$ contains just the vertex $B$.
It is easy to see that $|S_i|=|S_{-i}|=2^{d-i}$.
We define $\ket{\psi_i}=\frac{1}{\sqrt{|S_i|}} \sum_{v\in S_i}\ket{v}$.
Let $\H$ be the subspace spanned by states
$\ket{\psi_i}$, $i\in\{-d, -d+1, \ldots, d\}$.
Then, it can be shown\cite{CFG} that Hamiltonian $H$ maps $\H$ to itself.
This means that $e^{iHt}$ also maps $\H$ to itself.

Moreover, $H$ maps each $\ket{\psi_i}$ to a superposition of
$\ket{\psi_{i-1}}$, $\ket{\psi_i}$ and $\ket{\psi_{i+1}}$.
(For each $v\in S_i$, $H\ket{v}$ includes only $v$ and 
its neighbours. If $v\in S_i$, then the neighbours of $v$
belong to $S_{i-1}$ and $S_{i+1}$.)
Thus, we have a variant of continous quantum walk on line,
with matrix $H$ slightly different from one described
in section \ref{sec:cont}.
It can be analyzed using similar methods \cite{CFG,CC+}.
The result is that, if we start the walk in state $\ket{A}$
run for $t=O(d)$ steps and measure the state, then
the probability of measurement giving the state
$\ket{B}$ is $\Omega(\frac{1}{d})$.\cite{CC+} 
Repeating this process $d$ times finds $B$ with
a constant probability in time $O(d^2)$.

Essentially, this is an effect similar
to one described for discrete walk on the line in 
section \ref{sec:disc}. After $O(t)$ time, the walk
is in a superposition where $c*t$ locations have probability
$\Omega(\frac{1}{t})$. This effect is true for both
continous and discrete walk.

To summarize, classical random walk takes $O(2^d)$ steps to find
$B$ from $A$ but quantum walk of  \cite{CFG,CC+} takes
$O(d^2)$ steps. To claim an exponential speedup, 
it is also necessary to show that 
any classical algorithm (not just classical random walk)
requires exponentially larger 
number of steps. 
This was shown in  \cite{CC+} for a slight modification
of the above graph.

An important open problem is how to use this 
effect to design a quantum algorithm for
a natural problem. Problems involving Cayley graphs of groups might be good
candidates because those graphs have quite regular structure.

\section{Quantum walk search}
\label{sub:qws}

The algorithms of this type include refs. \cite{Shenvi,CG,ADistinct,AKRS}.
The basic situation is as follows. We have a graph $G$ in which
some vertices are marked. In one step, we are allowed to query 
if the current vertex is marked or move to an adjacent
vertex. Let $N$ be the number of vertices and $M$ be the longest
distance between two vertices. Then, Grover's algorithm \cite{Grover}
finds a marked vertex by querying just $O(\sqrt{N})$ vertices.

The problem is that, for any two vertices $v, v'$, there is a possibility that
$v$ is queried in one query and $v'$ is queried in the next query.
Thus, between each two queries, the algorithm might have to move
distance $M$. This results in the total running time $O(\sqrt{N} M)$.

While $\Omega(\sqrt{N})$ queries are required
for search\cite{BBBV}, we can design better algorithms by decreasing the
number of moving steps. Quantum walks are useful for that in several
different cases.

For discrete time quantum walks, the basic algorithm is as follows. 
For each vertex, we define two 
``coin flip'' transformations $C_v$ and $C'_v$. In each step of
the quantum walk, we query if the current vertex is marked.
If it is not marked, we apply $C_v$, otherwise, we apply $C'_w$.
Then, we perform a shift $S$ in the same way as before 
(sections \ref{sec:line}, \ref{sec:gen}).

\subsection{Searching hypercube}

The first algorithm of this type was discovered by Shenvi 
et al. \cite{Shenvi} for searching the Boolean hypercube.
In the Boolean hypercube, we have $N=2^n$ vertices $v_x$ indexed by
$n$ bit strings. Two vertices $v_x$ and $v_y$ are connected by an edge
if the corresponding strings $x$ and $y$ differ in exactly one
place. The maximum distance between two vertices is $n=\log N$.
Thus, Grover's algorithm could search this graph in $O(\sqrt{N}\log N)$ 
steps. Shenvi et al. \cite{Shenvi} showed how to search it by quantum 
walk in $O(\sqrt{N})$ steps. 
Consider Hilbert space spanned by
$\ket{x}\ket{i}$, where $x\in\{0, 1\}^n$ corresponds to a vertex
of hypercube and $i\in\{1, \ldots, n\}$ corresponds to an edge. 
The algorithm of  \cite{Shenvi} is as follows. 
\begin{enumerate}
\item
Generate the starting state 
$\frac{1}{\sqrt{2^n n}} \sum_{x, i}\ket{x}\ket{i}$.
\item
Perform $O(\sqrt{N})$ steps, with each step consisting of
\begin{enumerate}
\item
Apply $D_n$ to $\ket{i}$ register if the vertex $x$ is not marked.
Apply $-I$ to $\ket{i}$ register if the vertex $x$ is marked.
\item
Apply shift $S$ defined by
\[ \ket{x}\ket{i}\rightarrow \ket{x^i}\ket{i} \]
where $x^i$ is the bit string obtained by flipping the
$i^{\rm th}$ bit in $x$.
\end{enumerate}
\end{enumerate}

If there is no marked items, then the algorithm stays in
the starting state $\frac{1}{\sqrt{2^n n}} \sum_{x, i}\ket{x}\ket{i}$.
If there is a single marked vertex $v_x$, then, running this algorithm
for $O(\sqrt{N})$ steps and measuring the state gives $\ket{x, i}$ for the
marked $x$ and some $i$ with high probability (cf. Shenvi et al.
\cite{Shenvi}). The analysis is by reducing the walk on
hypercube to the walk on line, somewhat similarly to the
result of Childs et al. \cite{CFG} described in the previous
section.

\subsection{Searching a grid}

A second example is if $N$ items are arranged on $\sqrt{N}\times\sqrt{N}$
grid. Then, finding a marked item by usual quantum search
takes $O(\sqrt{N}*\sqrt{N})=O(N)$ steps, resulting in no quantum 
speedup \cite{Benioff}.
For $d$-dimensional $N^{1/d}\times N^{1/d}\ldots N^{1/d}$ grid,
the usual quantum search takes $O(\sqrt{N}*N^{1/d})=
O(N^{\frac{1}{2}+\frac{1}{d}})$ steps.
Childs and Goldstone \cite{CG}
studied quantum search on grids by continous walk and 
Ambainis et al. \cite{AKRS} studied search by discrete walk.
Surprisingly, continous and discrete walks gave different results.
Continous quantum walk algorithm gave an $O(\sqrt{N})$ time
quantum algorithm for $d\geq 5$, an $O(\sqrt{N} \log N)$ time
algorithm for $d=4$ and no speedup in 2 or 3 dimensions \cite{CG}.
Discrete time quantum walk gave an $O(\sqrt{N})$ time
algorithm for $d\geq 3$ and $O(\sqrt{N}\log N)$ time
algorithm for $d=2$.\cite{AKRS}

Another surprising result was that discrete walk algorithm
was very sensitive to the choice of coin transformations
$C_v$ and $C'_v$. One natural choice of coin transformations
gave an $O(\sqrt{N}\log N)$ step search algorithm while 
other natural choices gave either no quantum 
speedup in less than 4 dimension or no speedup at all.\cite{AKRS}

\subsection{Element distinctness}

The third application of quantum walk search is element
distinctness. 

{\bf Element Distinctness.}
Given numbers $x_1, \ldots, x_N\in[M]$, are there $i, j\in[N]$, $i\neq j$ such
that $x_i=x_j$?

Classically, element distinctness requires $\Omega(N)$ queries.
Ambainis \cite{ADistinct} has constructed an $O(N^{2/3})$ query
quantum algorithm using ideas from quantum walks. 
(Previously, an $O(N^{3/4})$ query quantum algorithm was 
known \cite{Distinctness}. $O(N^{2/3})$ is optimal, as shown by
refs. \cite{Shi,ASmallRange}.)

The main idea is as follows. We have vertices $v_S$ corresponding
to sets $S\subseteq\{1, \ldots, N\}$. Two vertices $v_S$ and $v_T$ 
are connected by an edge if sets $S$ and $T$ differ in one variable.
A vertex is marked if $S$ contains $i, j$ such that $x_i=x_j$.
At each moment of time, we know $x_i$ for all $i\in S$.
This enables us:
\begin{enumerate}
\item
Check if the vertex is marked with no queries.
\item
Move to an adjacent vertex $v_T$ by querying just one variable 
$x_i$ for the only $i$ such that $i\in T$, $i\notin S$.
\end{enumerate}
Then, we define quantum walk on subsets $S$.
 \cite{ADistinct} shows that, if $x_1, \ldots, x_N$ are not
distinct, this walk finds a set $S$ containing $i, j: x_i=x_j$
in $O(N^{2/3})$ steps.

Unlike in the previous two examples, the formulation of the problem
does not involve an obvious graph on which to define quantum walk.
Constructing the graph is a part of solution to the problem,
not the problem itself. 

Extensions of this algorithm also solve several other 
problems.

{\bf Element $k$-distinctness.}
Given numbers $x_1, \ldots, x_N\in[M]$, are there $k$ distinct indices
$i_1, \ldots, i_k\in [N]$ such that
$x_{i_1}=x_{i_2}=\ldots=x_{i_k}$?

Element $k$-distinctness can be solved in $O(N^{k/(k+1)})$ 
steps \cite{ADistinct}. It can be generalized to finding set of
$k$ element satisfying a certain property. This problem
can be also solved in $O(N^{k/(k+1)})$ quantum steps \cite{CE}.

{\bf Triangle finding.}
We are given a graph $G$ on $n$ vertices, described by variables 
$v_{ij}$, $v_{ij}=1$ if there is an edge between vertex $i$ and $j$
and $v_{ij}=0$ otherwise. Determine if graph $G$ contains
a triangle (vertices $i, j, k$ such that $v_{ij}=v_{ik}=v_{jk}=1$).

Triangle finding can be solved with $O(n^{1.3})$ queries \cite{MSS}
by a quantum algorithm that uses element distinctness as a subroutine. 
Similar approach also gives quantum algorithms for 
finding $k$-cliques (sets of $k$ vertices with an edge between
every two of them) \cite{MSS,CE}.

\section{Conclusion and open problems}
\label{sec:open}

Quantum walks provide a promising source of ideas for new quantum
algorithms. Several quantum algorithms using ideas from quantum 
walks have already been developed. In this paper, we surveyed
those algorithms and some of techniques used to prove their
correctness.

Some open problems are:
\begin{enumerate}
\item
{\bf Exponential speedup for natural problems.}
Can we use the exponentially faster hitting discovered by refs.
\cite{FG,CFG,CC+}
to get speedups for natural problems?
\item
{\bf Relation between discrete and continous quantum walks.}
Can discrete and continous quantum walks can be obtained one from
another in some way? This is the case for classical random walks.
However, in the quantum case, the relation between discrete and
continous walks is unclear, because one of them uses coin registers
and the other does not.
\item
{\bf Improving $k$-distinctness, triangle and $k$-clique algorithms.}
The $O(N^{2/3})$ element distinctness algorithm is known to be 
optimal \cite{Shi}. However, the best lower bound for $k$-distinctness
is only $\Omega(N^{2/3})$ \cite{ADistinct} and the best lower
bounds for triangles and $k$-cliques are $\Omega(n)$. Thus, better
quantum algorithms might be possible for all of those problems.
\item
{\bf Other applications for quantum walk search.}
In section \ref{sub:qws}, we showed how to solve three problems
in the same framework of searching a graph. It seems likely that
there might be other problems which can be described as 
search on graphs. Then, quantum walks might be
applicable to those problems as well.
\end{enumerate}

{\bf Acknowledgements.}
I would like to thank Viv Kendon and Alexander Rivosh for useful comments.

\end{document}